# Metastatic Cancer Image Classification Based On Deep Learning Method

GuanWen Qiu[1],
University of California, Irvine,
California, United States,
guanwenq@uci.edu,
Xiaobing Yu[1],
Miami University,
Ohio, United States,
yuio13912368355@gmail.com ,
Baolin Sun[2],
NorthEastern University,
Shenyang, China,
sunbaolin1994@163.com,

Yunpeng Wang[3],
University of Illinois, Urbana Champaign,
Illinois, United States,
yunpeng4@illinois.edu.
Lipei Zhang*
University College London
London, United Kindom
uceclz0@ucl.ac.uk

**Abstract—Using histopathological images to automatically classify cancer is a difficult task for accurately detecting cancer, especially to identify metastatic cancer in small image patches obtained from larger digital pathology scans. Computer diagnosis technology has attracted wide attention from researchers. In this paper, we propose a noval method which combines the deep learning algorithm in image classification, the DenseNet169 framework and Rectified Adam optimization algorithm. The connectivity pattern of DenseNet is direct connections from any layer to all consecutive layers, which can effectively improve the information flow between different layers. With the fact that RAdam is not easy to fall into a local optimal solution, and it can converge quickly in model training. The experimental results shows that our model achieves superior performance over the other classical convolutional neural networks approaches, such as Vgg19, Resnet34, Resnet50. In particular, the Auc-Roc score of our DenseNet169 model is 1.77% higher than Vgg19 model, and the Accuracy score is 1.50% higher. Moreover, we also study the relationship between loss value and batches processed during the training stage and validation stage, and obtain some important and interesting findings.**

## I. INTRODUCTION

Accurate metastatic cancer recognition is an essential task in medical diagnostic. In order to maximize the knowledge of cancer detection and interpretation, pathologists must study a large number of cancer tissue slides. Manual detection is expensive and has a certain error rate. Using histopathological images to automatically classify cancer is a difficult task for accurately detecting cancer, especially to identify metastatic cancer in small image patches obtained from larger digital pathology scans. With the rapid development of science and technology, computer diagnosis technology has attracted wide attention from researchers. Computer diagnosis is a reliable method of assessment and inspection, which can promote disease detection and reduce unnecessary manual diagnosis costs.

Traditional machine learning methods are labor intensive because they need manual extraction of visual features and handcrafted pre-processing before classification ([13], [14]). In recent years, deep learning methods can accurately, automatically, and sensitively extract features from images. Therefore, computer analysis based on deep learning has shown excellent benefits as a diagnostic strategy. Many efforts ([11], [12]) have been made to extend deep convolutional networks for metastatic cancer image classification which significantly improve diagnosis efficiency and accuracy.

In this paper, we propose utilize DenseNet Block with RAdam optimization algorithm to identify metastatic cancer in small image patches taken from larger digital pathology scans. We deal with the metastatic cancer image diagnosis as a binary image classification task in computer vision. For example, Figure 1 shows six small image patches taken from larger digital pathology scans images from Pcam dataset, where lable 1 is a cancer sample and the lable 0 is not a cancer sample. Our architecture is an automated framework that can be used to replace manual cancer detection and classification with high accuracy. Dense Convolutional Network (DenseNet) [1] has dense connectivity which can effectively capture the important imformation potential in images. The different connectivity pattern of our DenseNet with other convolutional neural networks is direct connections from any layer to all consecutive layers, which can effectively improve the information flow between different layers.

Moreover, we apply a novel optimization algorithm, Rectified Adam (RAdam) [2] which is effective and robust for model training. The optimizer of choice in many applications, Adagrad and its variants, e.g., RMSprop [5], Adam [4], and Nadam [16], stand out for their fast convergence. However, they usually obtain local optimal solutions. The training speed of SGD [3] algorithm is very slow, although it can converge to better results. However, RAdam is not easy to fall into a local optimal solution, and it can converge quickly. We conduct experiments on our dataset, which is the slightly modified version of the Pcam benchmark dataset provided by Kaggle competition. The experiments show that our model outperformed the other convolutional neural networks, such as Vgg19, Resnet34, Resnet50. The Auc-Roc score of our DenseNet169 model is 1.77% higher than Vgg19 model, and the Accuracy score is 1.50% higher. And we also study the relationship between loss value and batches processed in the training stage and validation stage, and obtain some interesting and important findings.

Our contributions are summarized as follows:
● This work is a new applied innovation to identify the metastatic cancer through DenseNet Block with RAdam optimization algorithm.We deal with the problem of metastatic cancer diagnosis as an image classification problem.
- We use RAdam optimization algorithm for model training, which is effective and robust.
- We evaluate our DenseNet169 method on the Pcam dataset. Experiments show that our model outperformed the comparison approaches, like Resnet34, Resnet50, and Vgg19. The relationship between loss value and batches processed in the training and validation stage indicates some important and interesting conclusions.

The rest of this article is laid out as follows. Section II introduces some latest research progress and related work about the metastatic cancer image classification and optimization algorithms. Next, Section III shows our proposed model based on a deep neural network, DenseNet Block. Then, we perform image classification experiments between our model and other classic deep neural network models in Section IV. Finally, Section V shows the conclusion of this paper and future work.

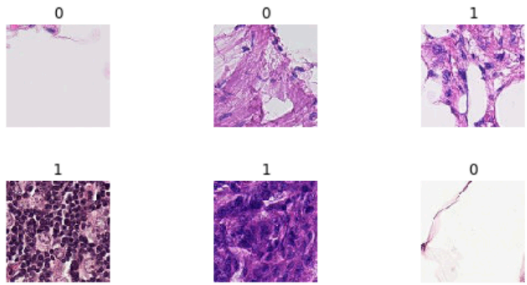

Figure1. Six small metastatic cancer images taken from larger digital pathology scans images from Pcam dataset. The lable 1 is a cancer sample and the lable 0 is not a cancer sample

## II. RELATED WORK

Accurate metastatic cancer recognizition is an essential task in medical diagnostic. With the rapid development of science and technology, computerized diagnosis has attracted a lot of attention from researchers. Computerized diagnosis is a great method of reliable evaluation and examinations, which can reduce unnecessary manual diagnosis costs and promote disease detection and classification. In recent years, computerized analysis based on deep learning has shown potential benefits as a diagnostic strategy. Application of deep learning algorithms to solve the metastatic cancer images classification task can significantly improve diagnostic efficiency and accuracy.

Traditional machine learning approaches ([13], [14]) require a lot of handcrafted pre-processing and manual extraction of visual features before classification, which are labor-intensive. But, deep learning methods can automatically, accurately and sensitively extract features from medical image data. Many efforts have been made to extend deep convolutional networks for medical image classification. For example, Wang et al. [15] used VGG16 deep neural network to metastatic breast cancer recognizition task. Habibzadeh et al. [12] conducted a study on breast cancer histopathological image classification task using ResNet model. And Veeling et al. [11] apply rotation equivariant CNNs for Digital Pathology.

For the optimization algorithm, in this paper, we use RAdam optimization algorithm [2]. Adagrad and its variants, e.g., RMSprop [5], Adam [4], and Nadam [16], stand out for their fast convergence, and have been regarded as the optimizer of choice in many applications. Although Adam and RMSProp converge very quickly, they usually obtain local optimal solutions. The early SGD [3] algorithm can converge to better results, but the training speed is very slow. RAdam is a new type of fast optimization algorithm, which has the advantages of both SGD algorithm and Adam algorithm. In addition, RAdam can converge quickly, and it is not easy to fall into a local optimal solution.

The aim of this study is to identify metastatic cancer in small image patches taken from larger digital pathology scans. Our proposed method use deep learning structure, DenseNet, with RAdam optimization algorithm which is helpful for model training. This work is a new application innovation to metastatic cancer classification. Our method is an automated architecture framework which is effective for cancer classification task.

## III. Methodology

Dense Convolutional Network (DenseNet) [3] can alleviate the vanishing-gradient problem, reduces lots of parameters, and enhance the propagation of feature maps. Because DenseNet has dense connectivity compared to other models like Resnet [6] and Vgg [7]. The different connectivity pattern of our DenseNet with other CNN is direct connections from any layer to all consecutive layers, which can effectively improve the information flow between different layers.

Some existing traditional convolutional neural networks connect the output of the $l$-th layer as input to the $(l+1)$-th layer [10], which is the following layer transition:

$$x^l = H^l(x^{l-1}). \qquad (1)$$

In particular, the ResNet architecture puts a skip connection, which uses an identity function to bypass the non-linear transformation. The formula is computed as follows:

$$x^l = H^l(x^{l-1}) + x^{l-1}, \qquad (2)$$

where $l$ means the layer, and $x^l$ indexes the output of the $l$-th layer. For each image sample, the dimension of input matrix is set to n × m.

An important difference between DenseNet and existing network methods is that DenseNet has very narrow layers. We assume that $H^l$ is a composite function of three consecutive operations: batch normalization (BN) [8], followed by a rectified linear unit (ReLU) [9] and a convolution (Conv). When each $H^l$ function produces $k$ feature maps, it follows that the $l$-th layer has $k_0 + k \times (l-1)$ input feature maps, where $k_0$ is the number of channels in the input layer. Figure 2 shows the overall structure of the DenseNet Block model. As shown in Figure 2, the $l$-th layer accepts the feature maps of all previous layers, the formula is defined as follows:

$$x^l = H^l([x^0, x^1, \ldots, x^{l-1}]), \qquad (3)$$

where $[x^0, x^1, \ldots, x^{l-1}]$ represents the concatenation of the feature maps generated at layers $0, 1, 2, \ldots, l-1$.

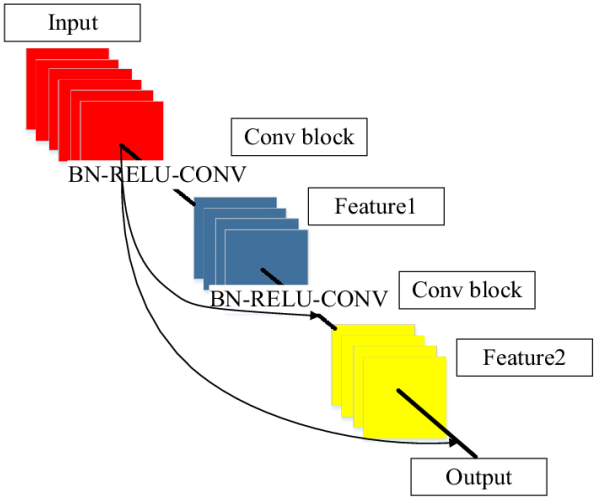

Figure2. The structure of DenseNet model.

## IV. Experiments

### A. Experimental Data and Experimental Settings

We conduct experiments between other classical convolutional neural network models and our proposed method for cancer image classification on our dataset in this section. First, we introduce the experimental data and experimental settings. We use the slightly modified version of PCam benchmark dataset to train and test our model which is provided by Kaggle competition. The PCam dataset takes the clinically relevant task of metastasis cancer detection into a straight forward binary image classification task. The initial PCam dataset includes repeated image samples due to its probabilistic sampling, but the modified version does not include repeated image samples. Therefore, the slightly modified PCam dataset has a total of 220025 images, including 130908 negative samples (without cancer, i.e. the label is 0) and 89117 positive samples (with cancer, i.e. the label is 1).

In the training stage and testing stage, our dataset is randomly split according to the ratio that training: test is 8: 2. In other words, four folds are used for training, and the testing dataset uses one fold. As shown in Section III, we minimize the loss function by using the RAdam optimizer [2] in our model during the training stage. The learning ratio is set to 0.0001 and the batch size is set to 128. All models which mentioned in this paper are trained with Pytorch on the GPU.

We apply two types of metrics: Auc-Roc score and Accuracy score to evaluate the performance of our method based DenseNet169 and RAdam. The Accuracy score is the sum of FP, TN, TP, and FN, where FP means the number of negative samples which are falsely classified. TN represents the number of negative samples which are truly classified, and TraditionalTP denotes the number of positive samples which are correctly predicted, FN is the number of positive samples which are incorrectly predicted. The Auc-Roc score metric is actually the area under roc curve, also known as the receiver operating characteristic curve. It can provide a comprehensive evaluation of model performance. The higher score means the model is better. The roc curve is drawn by plotting the TPR (true positive rate) and the FPR (false positive rate) at various threshold settings. Specifically, the formulas of FPR and TPR are calculated as follows:

$$FPR = \frac{FP}{FP+TN}, \qquad (4)$$

$$TPR = \frac{TP}{TP+FN}. \qquad (5)$$

### B. Experimental Results and Analyses

For our metastatic cancer image classification task, the compared models are Resnet34, Resnet50, and Vgg19. To evaluate the performance of our proposed model consistently, we compare these models on the same dataset and evaluation metrics. In Table 1, it shows the experimental results of different models on the Pcam dataset. It is easy to see that our DenseNet169 based model outperforms the other three models on both Auc Roc score and Accuracy score. As shown in Table 1, the DenseNet169 model with RAdam optimization algorithm has the highest score in both Auc-Roc metric and Accuracy metric. Especially, the Auc-Roc score of our DenseNet169 model is 1.77% higher than Vgg19 model, and the Accuracy score is 1.50% higher. Therefore, the experimental results demonstrate that our method is significant effective than other models for metastatic cancer image classification task.

| Models | Auc Roc Score | Accuracy |
|---|---|---|
| Resnet34 | 0.9633 | 0.975 |
| Resnet50 | 0.9642 | 0.976 |
| Vgg19 | 0.9473 | 0.965 |
| **Densenet169** | 0.965 | 0.980 |

Table 1 The results of different models on Pcam datasets for cancer image classification.

The line chart in Figure 3 presents the relationship between the loss value and the batches processed during the training stage and validation stage. We find that when the batches processed increases in multiples, the loss value of training stage and validation stage both keep decreasing. However, in the beginning, the loss value decrease obviously in both training stage and validation stage with the increase of batches processed. When the batches processed becomes very large (for example, batches processed ≥4000), the downward trend of the loss value becomes gradually gentle. It might be because that the model performance becomes better when the number of iterative training of the model is increased. But the model performance tends to be stable after reaching a certain batches processed. So, more iterative training may not guarantee better results, but it will waste time and computing resources. Moreover, the training loss value is higher than the verification loss value before a certain process moment. Correspondingly, the training loss value is lower than the verification loss value after the certain process moment. The reason maybe that when training to a certain level, the prediction performance of the model is better, rather than the lower training loss indicates the better the prediction performance of the model. Hence, above results could be constructive in guiding the future work direction for model training and verification. The proper batches processed is great for model training and verification.

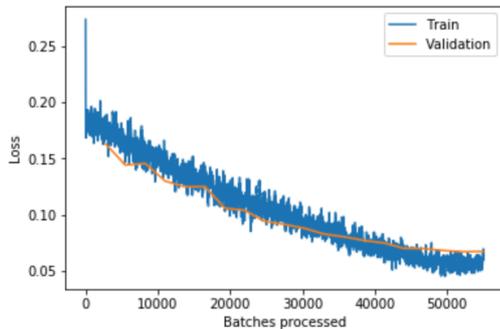

Figure 3. The relationship between loss value and batches processed in the training stage and validation stage.

## V. Conclusions

In this paper, we propose a novel DenseNet based model using RAdam optimization for metastatic cancer image classification task. This work is a new application innovation to identify the cancer. Our model employ RAdam optimization algorithm which is not easy to fall into the local optimal solution, and it can guarantee fast convergence than existing optimization algorithms. In addition, the DenseNet Block can effectively capture the important features of metastatic cancer images. The experiments on Pcam dataset indicate that our model outperform the comparison models, such as Resnet34, Resnet50, Vgg19. Our model has the highest Accuracy score and Auc Roc score than compared models. And the experimental results shows that the proper batches processed is important for model training and verification. In the future, we will explore our DenseNet based model for other image classification task.